\documentclass[prc,nofootinbib,showpacs]{revtex4}

\usepackage{graphicx}
\usepackage[usenames,dvipsnames]{color}
\usepackage{amsmath,amssymb,bbold}
\usepackage{hyperref}
\usepackage{comment}

\begin{document}

\title{Electrical conductivity and Hall conductivity of hot and dense quark gluon plasma in a magnetic field:
a quasi particle approach}

\author{Arpan Das$^{1}$}
\email{arpan@prl.res.in}
\author{Hiranmaya Mishra$^{1}$}
\email{hm@prl.res.in}
\author{Ranjita K. Mohapatra${^2}$}
\email{ranjita@iitb.ac.in}

\affiliation{$^{1}$Theory Division, Physical Research Laboratory, 
Navrangpura, Ahmedabad 380 009, India}
\affiliation{$^{2}$ Department of Physics, Indian Institute of Technology Bombay, Mumbai, 400076, India}

\begin{abstract}
We estimate here the electrical and Hall conductivity using a quasiparticle approach for quark matter. We use a Boltzmann kinetic
approach in presence of external magnetic field. We confront the results of model calculations with Lattice QCD simulations for 
vanishing magnetic field. In general electrical conductivity decreases with magnetic field. The Hall conductivity on the other hand can show a 
non monotonic behaviour with magnetic field due to an intricate interplay of behaviour of relaxation time and strength of the magnetic field.
We argue for vanishing quark chemical potential Hall conductivity vanishes and quark gluon plasma with finite quark 
chemical potential can show Hall effect. Both electrical conductivity and Hall conductivity increases with increasing quark chemical potential.
\end{abstract}

\maketitle

\section{INTRODUCTION}

\label{intro}

Relativistic heavy ion collision experiments at Relativistic Heavy Ion Collider (RHIC) and Large Hadron Collider (LHC)
provides an opportunity for a comprehensive understanding of quantum chromodynamics (QCD) in nonperturbative regime,
specially for the phase diagram of strongly interacting matter.
Large number of experimental data as well as 
theoretical models, give ample evidence of the formation and subsequent evolution of a deconfined strongly interacting
matter, known as quark-gluon plasma (QGP). It is expected that strongly interacting QCD plasma 
achieves local thermal equilibrium within about a 1 fm time. It is of paramount importance to understand the physical 
properties of the strong interacting matter produced in heavy ion collisions. 
For a comprehensive and detailed understanding of the hot and dense QCD  medium, transport 
coefficients, e.g. shear viscosity, bulk viscosity, electrical conductivity are very important.
These coefficients enter as essential theoretical input for the hydrodynamical simulations which are important tools for interpreting
heavy ion collision data.
A small shear viscosity to entropy ratio ($\eta/s$) is consistent with the transverse momentum spectra
of the charged particles within the framework of dissipative hydrodynamical model
of quark gluon plasma \cite{HeinzSnellings2013,RomatschkeRomatschke,KSS}.
$\eta/s$ of the strongly interacting plasma also satisfies the lower bound of shear viscosity to
entropy ratio, $\eta/s=\frac{1}{4\pi}$ ,
obtained using gauge gravity duality (AdS/CFT correspondence)\cite{KSS}.
This apart bulk viscosity $\zeta$, also plays an important role  
in the relativistic dissipative hydrodynamics describing the evolution of quark gluon plasma
\cite{gavin1985,kajantie1985, DobadoTorres2012,sasakiRedlich2009,sasakiRedlich2010,KarschKharzeevTuchin2008,
FinazzoRougemont2015,WiranataPrakash2009,JeonYaffe1996}. It is well known that trace of the energy momentum tensor of 
conformal fluid vanishes even at the quantum level. QCD is not conformal in nature and $(\epsilon-3P)/T^4$ is a measure of departure from 
conformality. The bulk viscosity which encodes the conformal measure of the system has been estimated using Lattice 
QCD simulations. First principle Lattice Monte Carlo simulations shows a non monotonic
behaviour of $\zeta/s$ as well as $\eta/s$ near the  critical temperature $T_c$.  \cite{DobadoTorres2012,sasakiRedlich2009,
sasakiRedlich2010,KarschKharzeevTuchin2008,FinazzoRougemont2015,WiranataPrakash2009,JeonYaffe1996}.

Further, plausibility of a generation of strong magnetic field in non central heavy ion collision experiments
brings novel phenomenological aspects. The strength of the magnetic field so produced 
strongly depends upon the center of mass energy of the collision. 
In fact at RHIC energies the strength of the magnetic field is expected to be as large as $eB\sim$ few $m_{\pi}^2$ and at LHC energies it can go 
even higher, of the order of $eB\sim$ 15 $m_{\pi}^2$, at least in the initial stage \cite{mclerran2008,skokov}.
Although the strength of the magnetic field is large at the initial stage in these collisions, 
strength of the magnetic field is expected to decay with time. Decay of the electromagnetic field in 
the absence of conducting medium is fast, but it has been argued that in a conducting medium 
due to induced currents the electromagnetic field does not decay very rapidly. 
Due to finite electrical conductivity ($\sigma^{el}$) of the plasma, the strength of the 
magnetic field can be significantly large in QGP even at later 
stages. A large electrical conductivity of plasma allows the magnetic field to survive 
in the medium for a relatively large time scale, leading to the magnetic field affecting significantly 
the evolution of the strong interacting matter 
 \cite{MHD1,MHDajit,TuchinMHD,MoritzGreif,electricalcond1,electricalcond2,electricalcond3,electricalcond4,
electricalcond5,electricalcond6,electricalcond7,electricalcond8,electricalcond9,Cassing, electricalcond10,electricalcond11,electricalcond12,
electricalcond13,electricalcond15}. This apart non vanishing magnetic field 
along with the topologically non trivial
non abelian QCD vacuum can give rise to novel CP violating effects such as chiral magnetic effect
and chiral vortical effect, etc \cite{kharzeevbook}. 
To estimate various transport coefficients of QGP and subsequent hadronic medium different complementary approaches 
e.g. perturbative QCD, QCD inspired effective models etc. have been investigated
in literature \cite{MoritzGreif,electricalcond1,electricalcond2,electricalcond3,electricalcond4,
electricalcond5,electricalcond6,electricalcond7,electricalcond8,electricalcond9,Cassing, electricalcond10,electricalcond11,electricalcond12,
electricalcond13,electricalcond15,danicol2018,PrakashVenu,WiranataPrakash2012,
KapustaChakraborty2011,Toneev2010,Plumari2012,Gorenstein2008,Greiner2012,TiwariSrivastava2012,GhoshMajumder2013,Weise2015,GhoshSarkar2014,
WiranataKoch,WiranataPrakashChakrabarty2012,Wahba2010,Greiner2009,KadamHM2015,Kadam2015,Ghoshijmp2014,Demir2014,Ghosh2014,smash,bamps,
urqmd1,GURUHM2015, ranjitahm,amanhm1,amanhm2,arpanhm}.
Apart from the viscosity coefficients, the conductivities both electrical and thermal conductivity
also plays an important role in the hydrodynamical evolution of 
strongly interacting medium 
at non zero baryon densities \cite{danicol2014,Kapusta2012}. Recently the thermoelectric effect of the 
hadronic medium produced at the later stages of heavy ion collisions has been investigated within the 
framework of hadron resonance gas model \cite{arpanhm}.

  In the present work, we investigate electrical conductivity as well as Hall conductivity of 
  quark gluon plasma produced in heavy-ion collisions.
  The Hall effect in a conducting medium is the manifestation of the generation of an induced electric current 
  transverse to an electric field and magnetic field (perpendicular to the electric field).
  Collision geometry in heavy ion collisions can give rise to configuration where electric field and magnetic field
  are transverse to each other \cite{tuchin1,tuchin2}. Hence it is interesting to study the Hall effect 
  for the electrically conducting QCD medium produced in heavy ion collisions.
   Hall conductivity of QGP has been studied within 
  the framework of perturbative QCD \cite{feng2017}. Recently we have studied Hall 
conductivity of the hadronic medium within the framework of hadron resonance 
gas model \cite{hmhall}.
  In the present investigation, we estimate the electrical and Hall conductivity for the 
  hot and dense QGP in a magnetic field using quasi particle picture of quark gluon plasma.
  It may be mentioned here that quasiparticle models have been quite successful in describing Lattice QCD simulation results 
  regarding thermodynamics \cite{plumari1,plumari2,plumari3,cassing1, quasiparticle1,
 quasiparticle2,quasiparticle3,quasiparticle4,quasiparticle5,quasiparticle6,quasiparticle7,quasiparticle8,quasiparticle9,
 quasiparticle10,quasiparticle11,quasiparticle12}. Generically in quasi particle picture non perturbative dynamics is encoded in the masses of quasiparticles. The 
 masses of the particles can be imagined to be arising from the energy contained in a strongly coupled volume 
 determined by the correlation range of the interaction. Once the effect of interaction is taken care of this way, the 
 quasi particles behave like free gas of massive constituents \cite{plumari1,plumari2,plumari3,cassing1, quasiparticle1}.
 Instead of medium dependent mass one can also consider the quasi particle 
 picture, by introducing effective fugacity parameter in the distribution function \cite{quasiparticle10}. This effective fugacity does not change the 
 mass of the particles rather it changes the single particle dispersion relation.
  It is important to note that in low energy condensed matter systems with specific type of 
  charge carriers (either electrons or holes) e.g. semiconductors etc., shows Hall effect \cite{semiconductor}. 
  Similarly an electron-ion plasma also shows Hall effect, because in electron-ion plasma mobility 
  of electrons and ions are different. Hence a net Hall current exists in electron-ion plasma. 
  However, for pair plasma (e.g. electron positron plasma) due to vanishing net gyration frequency 
  of the charge carriers net Hall current vanishes \cite{pairplasma1,pairplasma2,pairplasma3}. 
  Quark gluon plasma at vanishing baryon chemical potential is analogous to the case of pair plasma. 
  Hence QGP at vanishing quark chemical potential does not show Hall effect due to the exact cancellation 
  of Hall current due to particles and their antiparticles. 
  However at finite baryon chemical potential numbers of positive and negative charge carriers are not same, due to 
  asymmetry between the numbers of baryons and antibaryons. Hence electrically charged 
  quark gluon plasma with net baryon number, expected to be produced at low energy collisions 
  will also show Hall effect \cite{fairref,nica}.
  
  Keeping the above motivation in mind, we calculate the electrical conductivity and Hall conductivity 
  of quark gluon plasma in a magnetic field within the kinetic theory framework. 
  In Ref.\cite{feng2017} electrical conductivity and 
  Hall conductivity has been estimated using perturbative QCD approach. However QGP formed 
  in heavy ion collision experiments is 
  strongly coupled. Hence perturbative approach may not be sufficient to study the transport 
  coefficients of quark gluon plasma. In this context quasi particle model of QGP, 
  where quasi particle nature of the particles encode the nonperturbative effects, 
  have been used to study various transport coefficients
  \cite{cassing2,greco1,electricalcond3,patralata,lataguruhm2017,sukanyavinod,vinod2012,vinodmanu,vinodmanusuknya,manuvinod2018}. 
  In this investigation we also estimate the electrical conductivity and Hall
  conductivity in a magnetic field using quasi particle picture of QGP. 
  It is important to note that, although QGP medium can sustain  
  some fraction of the initial magnetic field due to finite electrical conductivity, 
  generically temperature is the dominant scale in the system.
  Hence in this case thermalization is governed by the strong interaction and the phase space and the single particle energies are not affected by 
  magnetic field through Landau quantization \cite{feng2017}.
  Effect of magnetic field only enters in the calculation through the cyclotron
  frequency of the charged particles. 
  Also in this work we have considered relaxation time approximation of the Boltzmann equation
  where external fields takes the system slightly away from equilibrium. This approximation is valid in this case because 
  thermalization is achieved due to the strong interaction and the external fields are generically 
  relatively small with respect to the dominant scale of the system. Hence external field only produces 
  a small deviation of the system from the equilibrium.

 This paper is organized as follows, in Sec. \ref{formalism} we briefly summarize the formalism to estimate electrical conductivity 
 and Hall conductivity using kinetic theory within relaxation time approximation as given in Ref.\cite{feng2017,hmhall}. 
 In Sec. \ref{quasiparticlemodel} we briefly discuss the quasi particle models of quark gluon plasma considered in this work.
 In Sec. \ref{results} we present and discuss the results for electrical and Hall conductivity. Finally we summarize our work with an outlook in the
 conclusion section.

\section{Boltzmann equation in relaxation time approximation}
\label{formalism}
In the presence of external electromagnetic field the relativistic Boltzmann transport equation (RBTE)
for a particle with electric charge $e$,  can be written as \cite{feng2017}, 

\begin{align}
 p^{\mu}\partial_{\mu}f(x,p)+eF^{\mu\nu}p_{\nu}\frac{\partial f(x,p)}{\partial p^{\mu}} = \mathcal{C}[f],
 \label{equ1}
\end{align}
where the electromagnetic field strength tensor is denoted as $F^{\mu\nu}$. 
On the right hand side of Eq.\eqref{equ1}, $\mathcal{C}[f]$ represents the collision integral which in the relaxation time
approximation (RTA) can be written as, 

\begin{align}
 \mathcal{C}[f]\simeq -\frac{p^{\mu}u_{\mu}}{\tau}(f-f_0)\equiv -\frac{p^{\mu}u_{\mu}}{\tau}\delta f ,
 \label{equ2}
\end{align}
where, $u_{\mu}$ is the fluid four velocity and in the local rest frame it has the form, $u_{\mu}\equiv(1,\vec{0})$.
In Eq.\eqref{equ2} $\tau$ is the thermal averaged relaxation time. Relaxation time  determines
the time scale over which a non equilibrium system relaxes towards its equilibrium state in the presence of small external perturbation.
The equilibrium state of the system is characterized by the equilibrium distribution function $f_0$. $f$ represents out of equilibrium distribution
function. The underlying assumption of the relaxation time approximation is that the external perturbation takes the  
system slightly away from equilibrium and then it relaxes towards equilibrium, exponentially with a time scale $\tau$.
In this approximation the external perturbation, which in this case is external electromagnetic field, not 
dominant scale with respect to the characteristic scale of the thermal system in equilibrium.
Hence we are not considering the effect of Landau quantization on the phase space of the particles and in the 
scattering processes. The equilibrium distribution  function satisfies ($f_0$), 

\begin{align}
 \frac{\partial f_0}{\partial \vec{p}}=\vec{v}\frac{\partial f_0}{\partial \epsilon},
 ~~ \frac{\partial f_0}{\partial \epsilon}=-\beta f_0(1-f_0),
 ~~f_0 = \frac{1}{1+e^{\beta(\epsilon\pm\mu)}},
 \label{equ3}
\end{align}
 where $\epsilon(p)=\sqrt{\vec{p}^2+m^2}$ is the  single particle energy, $\mu$ 
 is the quark chemical potential and $\beta=1/T$, is the 
 inverse of temperature. Using Eq.\eqref{equ2}, the Boltzmann equation \eqref{equ1} can be written in the following manner
 \cite{feng2017,hmhall},
 
 \begin{align}
  \frac{\partial f}{\partial t}+\vec{v}.\frac{\partial f}{\partial\vec{r}}+e\bigg[\vec{E}+\vec{v}\times\vec{B}\bigg]
  .\frac{\partial f}{\partial\vec{p}}=-\nu(f-f_0),
  \label{equ4}
 \end{align}
 where $\nu=1/\tau$ is the inverse of relaxation time. In case of uniform and static  medium where
 $f$ and $f_0$ are independent of the time and space \cite{feng2017}, Eq.\eqref{equ4} simplifies to, 
 
 \begin{align}
  -e\bigg[\vec{E}+\vec{v}\times\vec{B}\bigg]\frac{\partial f}{\partial\vec{p}} & = \nu(f-f_0),
  \label{equ5}
 \end{align}
 Here without loss of generality, electric field and magnetic field transverse to each other can be chosen in the following way,
 $\vec{E}=E\hat{x}$ and $\vec{B}=B\hat{z}$. For this representation of $\vec{E}$ and and magnetic field $\vec{B}$,
 Eq.\eqref{equ5} can be recasted as, 

\begin{align}
 \bigg(\nu-eB\bigg(v_x\frac{\partial}{\partial p_y}-v_y\frac{\partial}{\partial p_x}\bigg)\bigg)f(p)=\nu f_0(p)
 -eE\frac{\partial}{\partial p_x}f_0(p). 
 \label{equ6}
\end{align}

Eq.\eqref{equ6} can be solved using the following ansatz of the out of equilibrium 
distribution function $f(p)$ \cite{feng2017,hmhall},

\begin{align}
 f(p)=f_0-\frac{1}{\nu}e\vec{E}.\frac{\partial f_0(p)}{\partial \vec{p}}-\vec{\Xi}.\frac{\partial f_0(p)}{\partial \vec{p}}.
 \label{equ7}
\end{align}

Using the ansatz given in Eq.\eqref{equ7} and the Boltzmann equation as given in Eq.\eqref{equ6}, one can solve for $\vec{\Xi}$. It 
can be shown that for the choice of magnetic field and electric field the components of $\vec{\Xi}$ which satisfies the Eq.\eqref{equ6}
are \cite{hmhall},

\begin{align}
 \Xi_x=-eE\frac{\omega_c^2}{\nu(\nu^2+\omega_c^2)}, ~~~\Xi_y=-eE\frac{\omega_c}{\omega_c^2+\nu^2}, ~~~\Xi_z=0, 
 \label{equ8}
\end{align}
where, $\omega_c=eB/\epsilon(p)$, with $p\equiv|\vec{p}|$, is the cyclotron frequency of the charged particle. 
Using Eq.\eqref{equ8}, the ansatz for the out of equilibrium distribution function $f(p)$ as given in the Eq.\eqref{equ7} 
can be shown to be \cite{hmhall},

\begin{align}
 f(p) & =f_0-eEv_x\left(\frac{\partial f_0}{\partial \epsilon}\right)\frac{\nu}{\nu^2+\omega_c^2}
 +eEv_y\left(\frac{\partial f_0}{\partial\epsilon}\right)\frac{\omega_c}{\omega_c^2+\nu^2
 }.
 \label{equ9}
\end{align}

Electric current can be defined in the following way \cite{feng2017}, 

\begin{align}
 j^i=e\int\frac{d^3p}{(2\pi)^3}v^i\delta f =\sigma^{ij}E_j=\sigma^{el}\delta^{ij}E_j+\sigma^H\epsilon^{ij}E_j,
 \label{equ10}
\end{align}
where $\epsilon_{ij}$ is the anti symmetric $2\times2$ unity tensor, with $\epsilon_{12}=-\epsilon_{21}=1$.
Using Eq.\eqref{equ9} and Eq.\eqref{equ10} the electrical and the Hall 
conductivity can be identified as \cite{feng2017,hmhall}, 

\begin{align}
 \sigma^{el}=e^2\int\frac{d^3p}{(2\pi)^3}v_x^2\left(-\frac{\partial f_0}{\partial \epsilon}\right)\frac{\nu}{\nu^2+\omega_c^2},
\label{equ11}
 \end{align}

\begin{align}
 \sigma^{H}=e^2\int\frac{d^3p}{(2\pi)^3}v_y^2\left(-\frac{\partial f_0}{\partial \epsilon}\right)\frac{\omega_c}{\nu^2+\omega_c^2}.
\label{equ12}
 \end{align}

For an isotropic system the electrical conductivity and the Hall conductivity can be expressed as, 

\begin{align}
 \sigma^{el}=\frac{e^2}{3T}\int\frac{d^3p}{(2\pi)^3}\frac{p^2}{\epsilon^2} \frac{\nu}{\nu^2+\omega_c^2}f_0(1-f_0)=
 \frac{e^2}{3T}\int\frac{d^3p}{(2\pi)^3}\frac{p^2}{\epsilon^2} \frac{1/\tau}{(1/\tau)^2+\omega_c^2}f_0(1-f_0),
 \label{equ13}
\end{align}

\begin{align}
 \sigma^{H}=\frac{e^2}{3T}\int\frac{d^3p}{(2\pi)^3}\frac{p^2}{\epsilon^2} \frac{\omega_c}{\nu^2+\omega_c^2}f_0(1-f_0)=
 \frac{e^2}{3T}\int\frac{d^3p}{(2\pi)^3}\frac{p^2}{\epsilon^2} \frac{\omega_c}{(1/\tau)^2+\omega_c^2}f_0(1-f_0).
 \label{equ14}
\end{align}
It is important to note that in the absence of magnetic field, Eq.\eqref{equ13} reproduces the standard expression  of 
electrical conductivity in relaxation time approximation \cite{electricalcond3,lataguruhm}. 
Electrical and Hall conductivity for a system of multiple charge particle species can be expressed as, 

\begin{align}
 \sigma^{el}=\sum_i
 \frac{e^2_i\tau_ig_i}{3T}\int\frac{d^3p}{(2\pi)^3}\frac{p^2}{\epsilon^2_i} \frac{1}{1+(\omega_{ci}\tau_i)^2}f_0(1-f_0),
 \label{equ15}
\end{align}

\begin{align}
 \sigma^{H}= \sum_i\frac{e^2_i\tau_ig_i}{3T}\int\frac{d^3p}{(2\pi)^3}\frac{p^2}{\epsilon^2_i} \frac{\omega_{ci}\tau_i}
 {1+(\omega_{ci}\tau_i)^2}f_0(1-f_0),
 \label{equ16}
\end{align}
where $e_i$, $\tau_i$, $g_i$ and $\omega_{ci}$ are electric charge, thermal averaged relaxation time, degeneracy factor
and cyclotron frequency of the $i$-th charged particle species respectively.
It is easy to see from Eq.\eqref{equ15} and  Eq.\eqref{equ16} that  particles and their anti particles contribute to the
electrical conductivity in a same manner and their behaviour is opposite
in case of Hall conductivity. Using quasi particle picture of quark gluon plasma one can get relaxation time ($\tau$), 
medium dependent mass ($m$) as well as the medium dependent dispersion relation. Once these quantities are known 
 using Eq.\eqref{equ15} and Eq.\eqref{equ16} electrical conductivity and the Hall conductivity can be estimated. 
 
 \section{Quasi particle model of quark gluon plasma}
 \label{quasiparticlemodel}
 To describe the thermal properties of QGP one uses QCD at finite temperature and baryon chemical potential.
 At very high temperature, due to asymptotic freedom, a system of quarks and gluons can be treated as 
 ideal gas. But at relatively low temperature, near $T_c$, non perturbative effects become important.
 In the non perturbative domain first principle Lattice calculations give reliable prediction about different thermal 
 properties
 of the system. However for phenomenological aspects one needs an effective description of QGP near $T\sim T_c$.
 In this context one can use quasi particle description of quarks and gluons in medium to 
 investigate thermal properties of QGP. Basic idea behind various quasi particle pictures of QGP is that,
 one can approximate 
 the thermodynamic properties of a system of strongly interacting quarks and gluon by a system of quasi quarks
 and quasi gluons, where the information about the interaction is encoded in the physical properties e.g. medium dependent
 mass, of the quasi particles. In literature various types of quasi particle models are discussed \cite{plumari1,plumari2,plumari3,cassing1,quasiparticle1,
 quasiparticle2,quasiparticle3,quasiparticle4,quasiparticle5,quasiparticle6,quasiparticle7,quasiparticle8,quasiparticle9,
 quasiparticle10,quasiparticle11,quasiparticle12}. In this investigation we have considered two quasi particle models,
 quasi particle model I (QPM I) and quasi particle model II (QPM II). In QPM I, quark gluon plasma is described by an
 ideal gas of quasiparticles having temperature-dependent mass arising from
 the interactions with the surrounding quarks and gluons in the medium\cite{quasiparticle3,quasiparticle4,quasiparticle5}.
 In QPM II, we consider the effective fugacity quasi particle model (EQPM), where the quasi particle nature is 
 implemented by modifying  the distribution function for free quarks and gluons
 by introducing an effective fugacity, which encodes information of the interaction \cite{quasiparticle10}. 
 \subsection{Quasi Particle Model I (QPM I)}
 In this quasiparticle model, effective mass of all the quark (antiquark) has both bare mass ($m_{0}$) as well as 
 thermal mass ($m_{th}$), which can be expressed as \cite{quasiparticle3,quasiparticle4,quasiparticle5},
 \begin{align}
  m^2=m_{0}^2+\sqrt{2}m_{0}m_{th}+m_{th}^2.
  \label{equ17}
 \end{align}
The thermal mass ($m_{th}$) arises due to the interaction of quarks and antiquarks with the other constituents of the plasma and 
can be expressed as \cite{thermalmass}, 
\begin{align}
 m_{th}^2(T,\mu)= g^2(T,\mu)T^2\frac{N_c^2-1}{8N_c}\left(1+\frac{\mu^2}{\pi^2T^2}\right),
 \label{equ18}
\end{align}
where $g^2(T,\mu)$ is the two loop QCD running coupling constant at finite temperature ($T$) and quark chemical
potential ($\mu$)\cite{quasiparticle3,quasiparticle5},
\begin{align}
g^2(T,\mu)=\frac{24\pi^2}{(33-2n_f)\ln\left(\frac{T}{\Lambda_T}\sqrt{1+a\frac{\mu^2}{T^2}}\right)}
\left(1-\frac{3(153-19n_f)}{(33-2n_f)^2}\frac{\ln\left(2\ln\left(\frac{T}{\Lambda_T}\sqrt{1+a\frac{\mu^2}{T^2}}\right)
\right)}{\ln\left(\frac{T}{\Lambda_T}\sqrt{1+a\frac{\mu^2}{T^2}}\right)}\right),
\label{equ19}
\end{align}
where $\Lambda_T$ is the QCD scale parameter and $a=1/\pi^2$.

The relaxation time $\tau$ of quarks (antiquarks) and gluon can be given by the following expression
\cite{kajantie1985,lataguruhm2017}, 

\begin{align}
 \tau_{q(\bar{q})}=\frac{1}{5.1T\alpha_s^2\ln(\frac{1}{\alpha_s})(1+0.12(2n_f+1))},
 \label{equ20}
\end{align}
and,
\begin{align}
 \tau_g=\frac{1}{22.5T\alpha_s^2\ln(\frac{1}{\alpha_s})(1+0.06n_f)},
 \label{equ21}
\end{align}
where, $\alpha_s(T,\mu)=\frac{g^2(T,\mu)}{4\pi}$. $g^2(T,\mu)$ is the temperature ($T$) and quark chemical potential ($\mu$) dependent
strong coupling constant. It is important to note  that the above mentioned relaxation time as given in Eq.\eqref{equ20}
and Eq.\eqref{equ21}, has been derived in massless case and for $\mu<<T$. 
It has been argued in Ref.\cite{massrelaxation} that the effect mass on the scattering cross sections is small. Thus mass
has non qualitative difference in the relaxation time as well as on transport coefficient in the quasi particle model. 
In this model light quark masses have been chosen to be $0.1$ times the strange quark mass, which is consistent with the 
chiral perturbation theory results \cite{chipt1,chipt2,chipt3}. The parameters $\Lambda_T/T_c$ and the strange quark mass
can be adjusted to fit the Lattice data\cite{chipt1}. The fitted parameters are $\Lambda_T/T_c = 0.35$, with $T_c=200$MeV 
and $m_{s0}= 80$ MeV\cite{quasiparticle5}.
 \subsection{Quasi Particle Model II (QPM II)}
\begin{figure}[!htp]
\begin{center}
\includegraphics[width=0.6\textwidth]{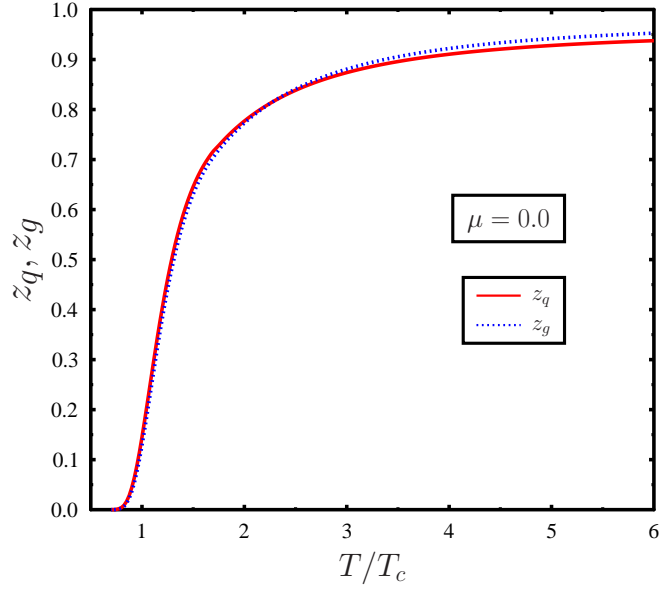}
\caption{Variation of $z_q$ and $z_g$ with temperature ($T$).}
\label{fig0}
\end{center}
\end{figure}

In this model the basic ansatz is that Lattice
QCD EOS can be reproduced in terms of non-interacting
quasi particle degrees of freedom having effective fugacities ($z_q,z_g$) which encodes
all the interaction effects of the particles in the system. These effective fugacities enters through the equilibrium 
distribution function of gluons and quarks (antiquarks) at finite temperature and vanishing baryon chemical potential
, which can be expressed as\cite{epqm1}, 

\begin{align}
 f^g_0=\frac{z_g\exp(-\beta p)}{1-z_g\exp(-\beta p)},
 \label{quasigludis}
\end{align}

\begin{align}
 f^{q/\bar{q}}_0=\frac{z_q\exp(-\beta \epsilon(p))}{1+z_q\exp(-\beta \epsilon(p))}
 =\frac{z_q\exp(-\beta \sqrt{p^2+m^2})}{1+z_q\exp(-\beta \sqrt{p^2+m^2})}.
 \label{quasiquarkdis}
\end{align}
From Eq.\eqref{quasigludis} and Eq.\eqref{quasiquarkdis} one gets the equilibrium distribution function 
of ideal quarks and gluon in the limit when quark/antiquark fugacity ($z_q$) and gluons fugacity ($z_g$) approaches unity,
i.e. $z_q\sim z_g \sim 1.0$.
For the complete information of the quasi particle distribution functions as given in Eq.\eqref{quasigludis} and 
Eq.\eqref{quasiquarkdis} one requires the temperature dependence of 
the fugacities $z_{q}$ and $z_g$. Temperature dependence of fugacities $z_{q}$ and $z_g$ can be 
obtained by matching the thermodynamic properties of this quasi particle model with that of Lattice QCD results.
Fitting the thermodynamic properties of this model with Lattice QCD 
data, for phenomenological purpose one can get a parametric form for $z_{q}$ and $z_g$ as a function of temperature \cite{epqm1}. 
Following Ref.\cite{epqm1}, the temperature dependence of the fugacities $z_{q}$ and $z_g$ can be given as,
 
\begin{align}
 & z_{q,g} = a_{q,g}\exp(-b_{q,g}/x^5), ~~\text{for},~~ x<x_{q,g},\nonumber\\
 & z_{q,g} = a^{\prime}_{q,g}\exp(-b^{\prime}_{q,g}/x^2), ~~\text{for},~~ x>x_{q,g},~~x_{q,g}\equiv 
 T_{q,g}/T_c\sim 1.70,1.68,
\end{align}
where the fitting parameters $a_{q,g}$, $a^{\prime}_{q,g}$, 
$b_{q,g}$, $b^{\prime}_{q,g}$ are given in  table\eqref{table1}\cite{epqm1}.
In this investigation for simplicity we have considered, $T_{q,g}/T_c = T/T_c \sim 1.70$ and the central values of 
all the fitting parameters for $z_{q}$ and $z_g$. Variation of $z_{q}$ and $z_g$ with temperature as shown in the 
Fig.\eqref{fig0} is consistent with the estimate of $z_{q}$ and $z_g$ as given in Ref.\cite{epqm1}.

\begin{table}
\centering
  \begin{tabular}{|c| c| c| c|c|}
   \hline
 $z_{g,q}$ & $a_{g,q}$ & $b_{g,q}$ & $a^{\prime}_{g,q}$ & $b^{\prime}_{g,q}$ \\ 
 \hline\hline
 Gluon & 0.803 $\pm$ 0.009 & 1.837 $\pm$ 0.039 & 0.978 $\pm$ 0.007 & 0.942 $\pm$ 0.035 \\ 
 Quark & 0.810 $\pm$ 0.010 & 1.721 $\pm$ 0.040 & 0.960 $\pm$ 0.007 & 0.846 $\pm$ 0.033 \\
 \hline
  \end{tabular}
 \caption{Fitting parameters for $z_{q}$ and $z_g$ \cite{epqm1}.}
 \label{table1}
\end{table}

It can be shown that effective fugacity parameters in the equilibrium distribution 
function  affect single particle dispersion relation in the following way\cite{epqm1},

\begin{align}
 & \omega_p^{g}=p+T^2\partial_T\ln(z_g)\\
 & \omega_p^{q}=\sqrt{p^2+m^2}+T^2\partial_T\ln(z_{q}).
 \end{align}
Although the dispersion relations are modified, it is important to note that in effective fugacity quasi particle model,
group velocity of the quasi particles remains unchanged, i.e.
\begin{align}
 \vec{v}=\partial_{\vec{p}}\omega_{p} = \partial_{\vec{p}}\epsilon(p) = \frac{\vec{p}}{\epsilon(p)}.
\end{align}

This model can be extended at finite quark chemical potential by introducing quark chemical potential in quark/antiquark distribution
function\cite{epqm2},

\begin{align}
  f^{q/\bar{q}}_0=\frac{z_q\exp(-\beta (\epsilon(p)\mp\mu))}{1+z_g\exp(-\beta (\epsilon(p)\mp\mu))}
 =\frac{z_q\exp(-\beta (\sqrt{p^2+m^2}\mp\mu))}{1+z_q\exp(-\beta (\sqrt{p^2+m^2}\mp\mu))}.
\end{align}
It is important to note that although quark chemical potential is introduced in the distribution function at finite 
chemical potential, fugacities are assumed to be independent of chemical potential as they are fixed with Lattice data 
at finite temperature and zero chemical potential. 

The thermal averaged relaxation time of quarks, antiquarks and gluons at finite temperature and chemical potential
in the effective fugacity quasi particle model has been considered as \cite{epqm2,chaozhang},

\begin{align}
 \tau^{-1}_g= & g_g\int \frac{d^3\vec{p}_g}{(2\pi)^3}f^g_0(1+f^g_0)\left(\frac{9g_{eff}^4}{16\pi\langle s\rangle_{gg}}
 \left[\ln\frac{\langle s\rangle_{gg}}{k^2}-1.267\right]\right)
  + g_q\int \frac{d^3\vec{p}_q}{(2\pi)^3}f^q_0(1-f^q_0)\left(\frac{g_{eff}^4}{4\pi\langle s\rangle_{gq}}
 \left[\ln\frac{\langle s\rangle_{gq}}{k^2}-1.287\right]\right)\nonumber\\
 &~~~~~~~~~~~~~~~~~~~~~~~~~ + g_{\bar{q}}\int \frac{d^3\vec{p}_{\bar{q}}}{(2\pi)^3}f^{\bar{q}}_0(1-f^{\bar{q}}_0)\left(\frac{g_{eff}^4}{4\pi\langle
 s\rangle_{g\bar{q}}}\left[\ln\frac{\langle s\rangle_{g\bar{q}}}{k^2}-1.287\right]\right),
 \label{equ29}
\end{align}

\begin{align}
 \tau^{-1}_q= & g_g\int \frac{d^3\vec{p}_g}{(2\pi)^3}f^g_0(1+f^g_0)\left(\frac{g_{eff}^4}{4\pi\langle s\rangle_{qg}}
 \left[\ln\frac{\langle s\rangle_{qg}}{k^2}-1.287\right]\right) + g_q\int \frac{d^3\vec{p}_q}{(2\pi)^3}f^q_0(1-f^q_0)\left(\frac{g_{eff}^4}{9\pi\langle s\rangle_{qq}}
 \left[\ln\frac{\langle s\rangle_{qq}}{k^2}-1.417\right]\right)\nonumber\\
 & ~~~~~~~~~~~~~~~~~~~~~~~~~~~~+ g_{\bar{q}}\int \frac{d^3\vec{p}_{\bar{q}}}{(2\pi)^3}f^{\bar{q}}_0(1-f^{\bar{q}}_0)\left(\frac{g_{eff}^4}{9\pi\langle
 s\rangle_{q\bar{q}}}\left[\ln\frac{\langle s\rangle_{q\bar{q}}}{k^2}-1.417\right]\right),
 \label{equ30}
\end{align}

\begin{align}
 \tau^{-1}_{\bar{q}}= & g_g\int \frac{d^3\vec{p}_g}{(2\pi)^3}f^g_0(1+f^g_0)\left(\frac{g_{eff}^4}
 {4\pi\langle s\rangle_{\bar{q}g}}
 \left[\ln\frac{\langle s\rangle_{\bar{q}g}}{k^2}-1.287\right]\right)+ g_q\int \frac{d^3\vec{p}_q}{(2\pi)^3}f^q_0(1-f^q_0)
 \left(\frac{g_{eff}^4}{9\pi\langle s\rangle_{\bar{q}q}}
 \left[\ln\frac{\langle s\rangle_{\bar{q}q}}{k^2}-1.417\right]\right)\nonumber\\
 &~~~~~~~~~~~~~~~~~~~~~~~~~~~~~+ g_{\bar{q}}\int \frac{d^3\vec{p}_{\bar{q}}}{(2\pi)^3}f^{\bar{q}}_0
 (1-f^{\bar{q}}_0)\left(\frac{g_{eff}^4}{9\pi\langle
 s\rangle_{\bar{q}\bar{q}}}\left[\ln\frac{\langle s\rangle_{\bar{q}\bar{q}}}{k^2}-1.417\right]\right),
\end{align}
where thermal average of the quantity $s$ is denoted as $\langle s \rangle_{kl}=2\langle p_k\rangle \langle p_l\rangle$, with 
$\langle p_k\rangle = \frac{\int \frac{d^3p_k}{(2\pi)^3}|\vec{p}_k|f^k_0}{\int \frac{d^3p_k}{(2\pi)^3}|f^k_0}$
and $k^2=g_{eff}^2T^2$.
Effective strong coupling constant $g_{eff}$ in this model can be  determined using charge renormalization, by
computing the Debye mass in the medium within the framework of effective fugacity quasi particle model and comparing
it to the hard thermal loop results. For $\mu/T\equiv\tilde{\mu}<1$, effective strong coupling constant
can be shown to be \cite{epqm2},

\begin{align}
 \alpha_{s_eff}(T,\mu)=\frac{g_{eff}^2}{4\pi}=\alpha_s(T,\mu)
 \frac{\frac{2N_c}{\pi^2}\text{PolyLog}[2,z_g]-\frac{2N_f}{\pi^2}\text{PolyLog}[2,-z_q]+\tilde{\mu}^2\left(\frac{N_f}{\pi^2}
 \frac{z_q}{1+z_q}\right)}{\left(\frac{N_c}{3}+\frac{N_f}{6}\right)+\tilde{\mu}^2\frac{N_f}{2\pi^2}},
\end{align}
where, $\alpha_s(T,\mu)$ is the temperature and chemical potential dependent strong coupling constant\cite{quasiparticle5}.
In this investigation we have considered strange quark mass ($m_s$) to be $80$ MeV and the light quark masses are taken as one tenth 
of the strange quark mass for QPM II. QCD transition temperature is taken as $T_c=200$MeV\cite{chipt1}.

\section{results and discussions}
\label{results}

\begin{figure}[!htp]
\begin{center}
\includegraphics[width=0.6\textwidth]{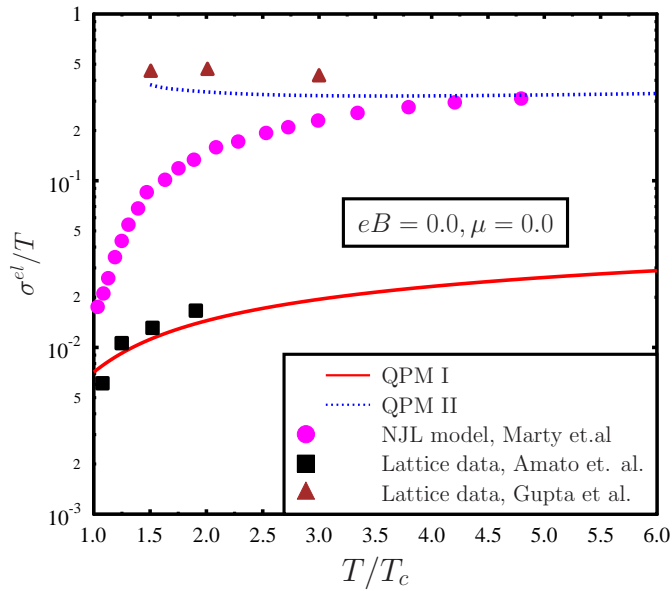}
\caption{Variation of normalized electrical conductivity ($\sigma^{el}/T$) with temperature ($T$) at vanishing
magnetic field and vanishing quark chemical potential. Lattice data are also shown in this plot for comparison.
Value of $\sigma^{el}/T$ in QPM I is consistent with Lattice data given by Amato 
et al.\cite{electricalcond8}. On the other hand $\sigma^{el}/T$ as estimated in QPM II is consistent with the 
Lattice QCD data given by Gupta et.al.\cite{electricalcond9}.}
\label{fig1}
\end{center}
\end{figure}

\begin{figure}[h]
\centering
\begin{minipage}{0.45\textwidth}
\centering
\includegraphics[width=1.1\linewidth]{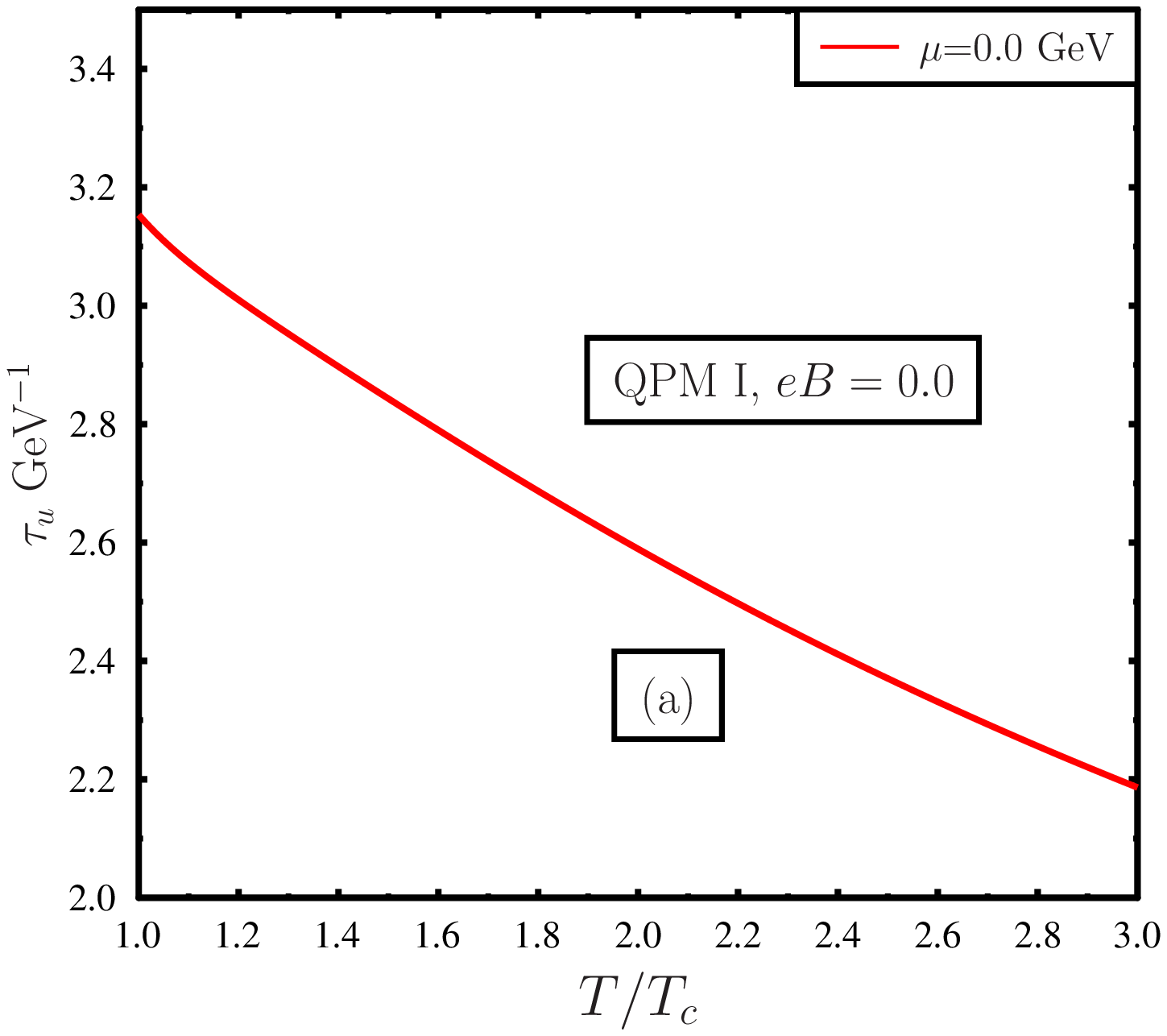}
\end{minipage}
\begin{minipage}{0.45\textwidth}
\centering
\includegraphics[width=1.1\linewidth]{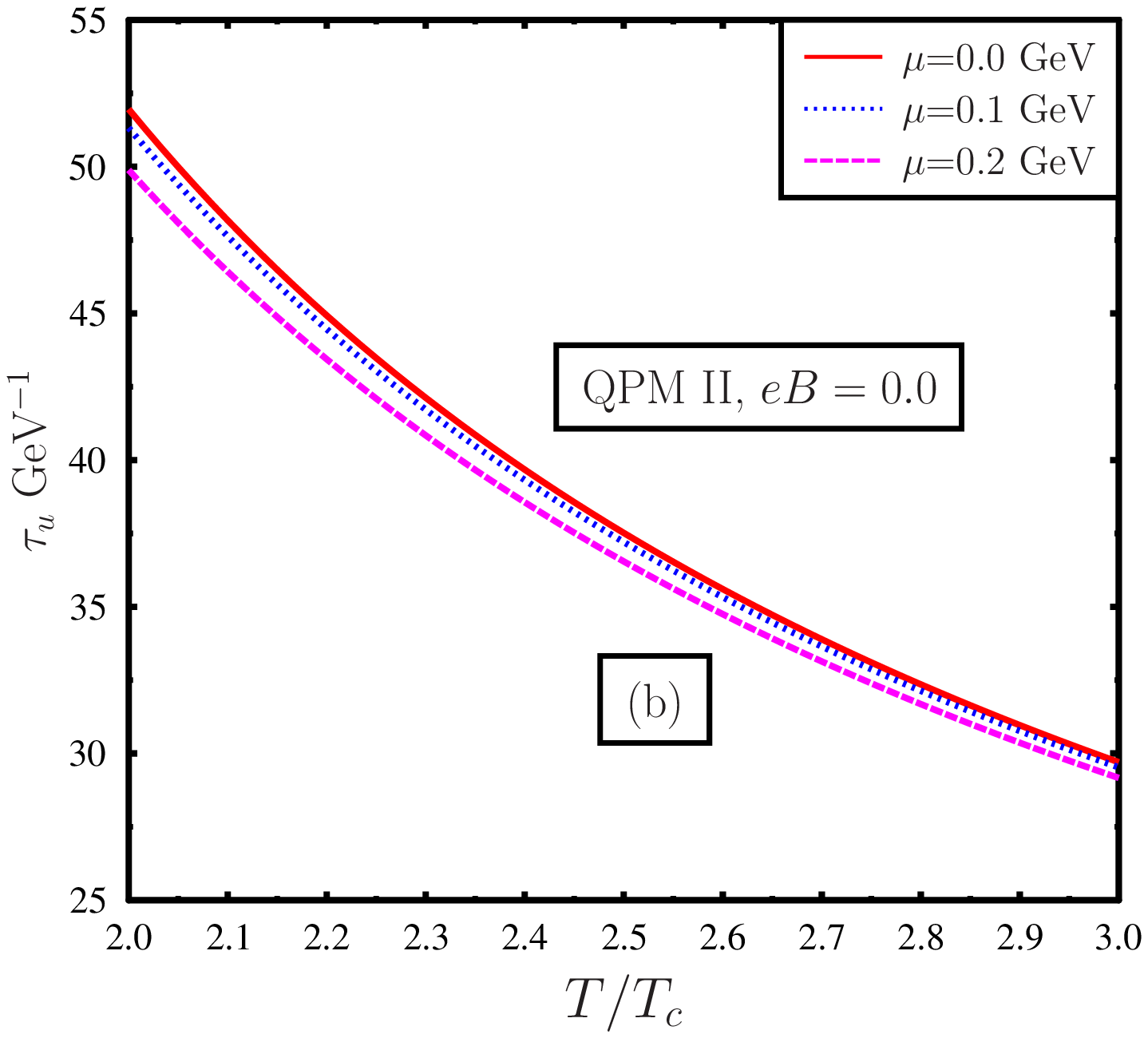}
\end{minipage}
\caption{Plot(a): Variation of thermal averaged relaxation time of $u$ quark with temperature for vanishing magnetic field and quark chemical
potential in QPM I. Plot(b): Variation of relaxation time of $u$ quark with temperature for 
vanishing magnetic field but with different values quark chemical potential in QPM II. 
From plot(a) and plot(b) we see that with increasing temperature thermalized relaxation time decrease for both the quasi particle
models.  For QPM II, relaxation time decreases with increase in quark chemical potential.}
\label{fig2}
\end{figure}

In Fig.\eqref{fig1} we show the variation of the normalized electrical conductivity ($\sigma^{el}/T$) with 
temperature,
at vanishing magnetic field ($eB=0.0$) and quark chemical potential ($\mu$) for QPM I and QPM II. From this figure it is clear that 
normalized electrical conductivity ($\sigma^{el}/T$) in QPM I is consistent with the Lattice QCD data given by Amato 
et al.\cite{electricalcond8}. On the other hand normalized electrical conductivity ($\sigma^{el}/T$) as estimated in QPM II is 
consistent with the Lattice QCD data given by Gupta et.al.\cite{electricalcond9}. For comparison we have also shown in Fig.\eqref{fig1},
normalized electrical conductivity obtained in Nambu-Jona-Lassinio model as given by Marty et.al.\cite{Cassing}. 
Value of the normalized electrical conductivity 
in these two quasi particle models are order of magnitude different. This is because electrical conductivity
is proportional to the relaxation time, which is order of magnitude larger in QPM II with respect to QPM I as 
can be seen from Fig.\eqref{fig2}.

In Fig.\eqref{fig2}(a) and in Fig.\eqref{fig2}(b) we show the variation of relaxation time of $u$ quarks with temperature ($T$)
and quark chemical potential ($\mu$). For the QPM I  relaxation time of the quarks and gluons as given in Eq.\eqref{equ20} and Eq.\eqref{equ21}
were derived for the case of $\mu<<T$\cite{kajantie1985}. Hence throughout this investigation we have considered relaxation 
time in QPM I for vanishing quark chemical potential. It is important to mention that in this investigation we have not considered the 
effect of magnetic field on the relaxation time as magnetic field is not the dominant scale. From Fig.\eqref{fig2} it is 
clear that the thermal average relaxation time in QPM II is order of magnitude larger that its counterpart in QPM I.
This is because of the different quasi particle nature of the partons in these models as can be seen from Eq.\eqref{equ20} and 
Eq.\eqref{equ30}. This apart from Fig.\eqref{fig2}(a)  and Fig.\eqref{fig2}(b) we can also see that with increasing temperature
relaxation time decreases. Physically this means with increasing temperature as the number density of the partons increases
collision rate increases. From Fig.\eqref{fig2}(b)
it is clear that with increasing chemical potential relaxation time decreases in QPM II. Although 
the dependence of relaxation time on the chemical potential in QPM II is convoluted as can be seen from Eq.\eqref{equ30},
but physically one can understand the variation of relaxation time with quark chemical potential in the following way.
With increasing chemical potential number density of the scatterer increases due to increasing number density 
of the particles. Hence interaction rate increases with increasing quark chemical potential, which gives rise to 
decreasing behaviour of relaxation time with increasing quark chemical potential.

\begin{figure}[h]
\centering
\begin{minipage}{0.45\textwidth}
\centering
\includegraphics[width=1.1\linewidth]{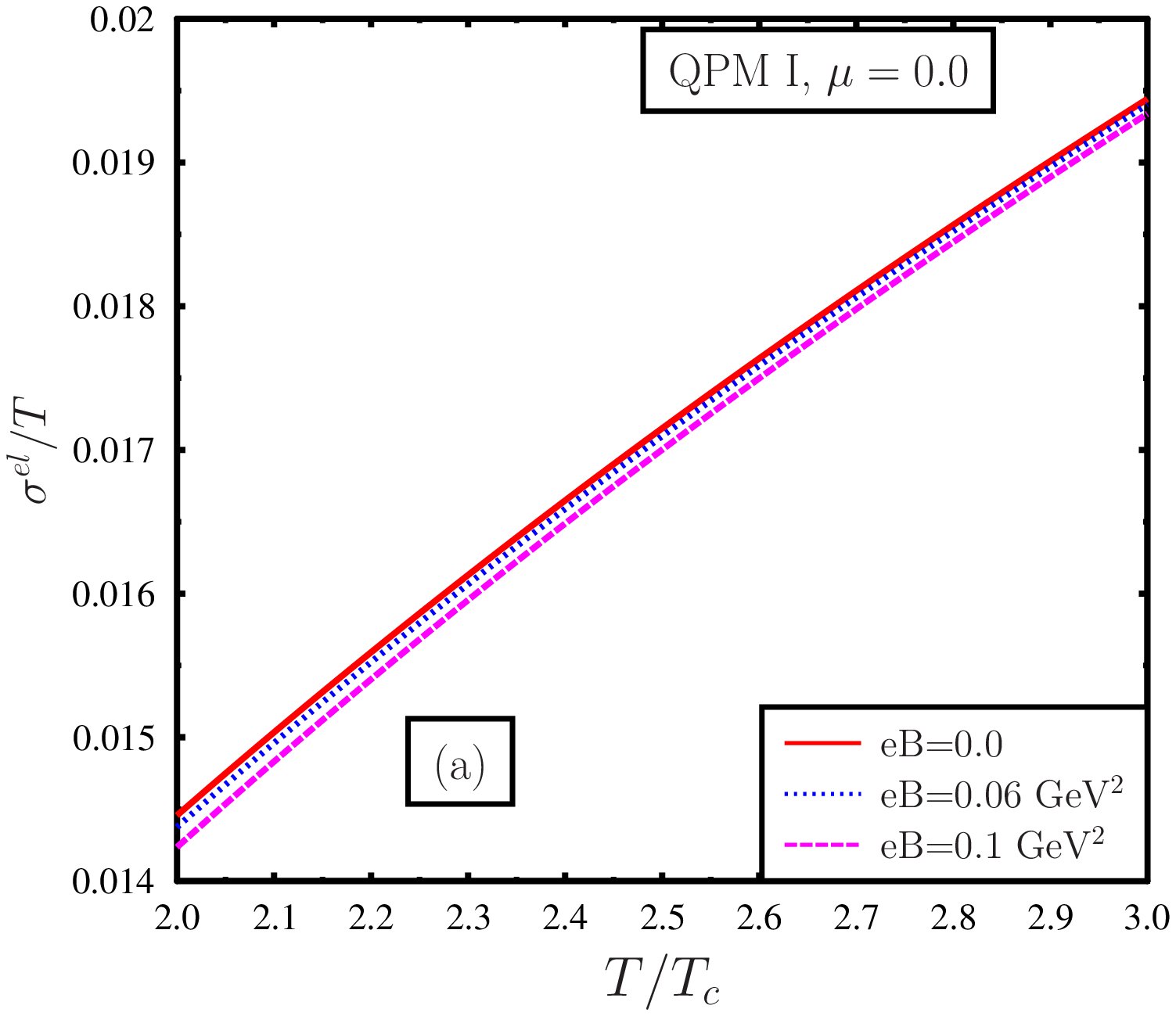}
\end{minipage}
\begin{minipage}{0.45\textwidth}
\centering
\includegraphics[width=1.1\linewidth]{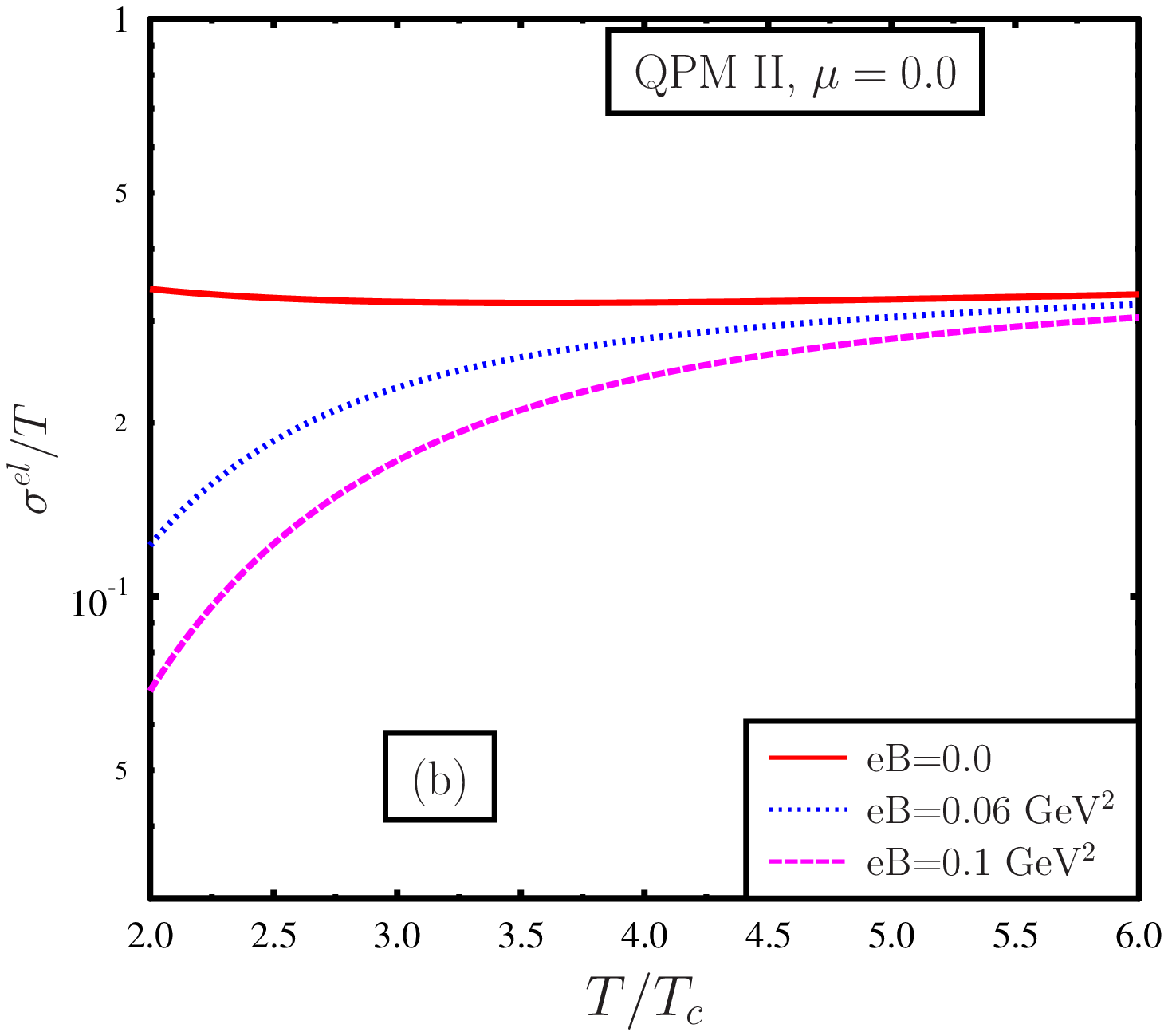}
\end{minipage}
\caption{Plot(a):Variation of normalized electrical conductivity ($\sigma^{el}/T$) with temperature for vanishing quark chemical
potential but with different values of magnetic fields in QPM I. Plot(b):Variation of normalized electrical conductivity
($\sigma^{el}/T$) with temperature for vanishing quark chemical
potential but with different values of magnetic fields in QPM II. From plot(a) and (b) we can see that for both the quasi particle 
models, normalized electrical conductivity decreases with increasing magnetic field. Decrease of normalized electrical 
conductivity is significantly larger in QPM II with respect to its counterpart in QPM I.}
\label{fig3}
\end{figure}

In Fig.\eqref{fig3} we show the variation of normalized electrical conductivity with temperature at vanishing quark chemical potential 
but with finite magnetic field. From Fig.\eqref{fig3}(a) and Fig.\eqref{fig3}(b), we see that with 
increasing magnetic field normalized electrical 
conductivity ($\sigma^{el}/T$) decreases. In QPM II the decrease in $\sigma^{el}/T$ is much larger than its 
counterpart in QPM I. This behaviour can be understood from Eq.\eqref{equ15}. Since in QPM I value of the 
relaxation time is order of magnitude smaller than the relaxation time in QPM II, $\omega_c\tau$ in the denominator in 
Eq.\eqref{equ15} is larger in QPM II. This gives rise to larger decrease in the normalized electrical conductivity
in QPM II with respect to QPM I. Physically electrical conductivity decreases with magnetic field because 
with increasing magnetic field more particles are deviated from the direction of electric field, hence reduction in electrical current. 
Variation of normalized electric conductivity with temperature for a fixed magnetic field and quark chemical potential
depends crucially on the temperature dependence of relaxation time and the equilibrium distribution function. 
For the range of quark chemical potential, temperature and magnetic field considered in this investigation,
from Fig.\eqref{fig3}(a) we can see that with temperature $\sigma^{el}/T$ increases. This increasing behaviour of 
$\sigma^{el}/T$ is predominately due to the Boltzmann factor $\exp(-\epsilon(p)/T)$ in the distribution function,
which increases with increasing temperature. Similarly for QPM II with temperature normalized electrical conductivity
increases for non vanishing magnetic field as can be seen in Fig.\eqref{fig3}(b).
With increasing temperature the relaxation time decreases and 
 the Boltzmann factor in the distribution function increases, giving rise to this increasing behavior of 
normalized electrical conductivity at non vanishing magnetic field in QPM II.

\begin{figure}[h]
\centering
\begin{minipage}{0.45\textwidth}
\centering
\includegraphics[width=1.1\linewidth]{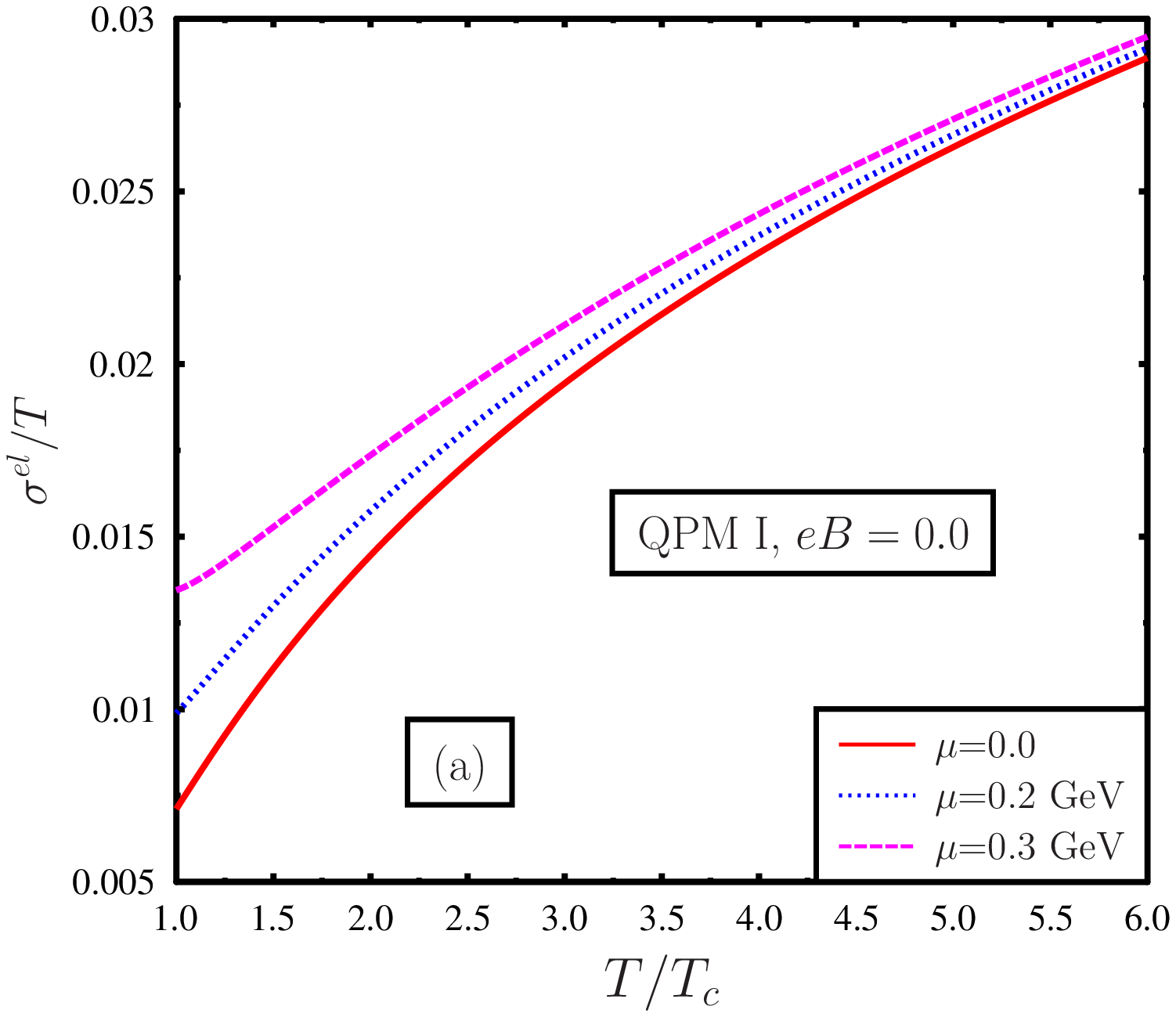}
\end{minipage}
\begin{minipage}{0.45\textwidth}
\centering
\includegraphics[width=1.1\linewidth]{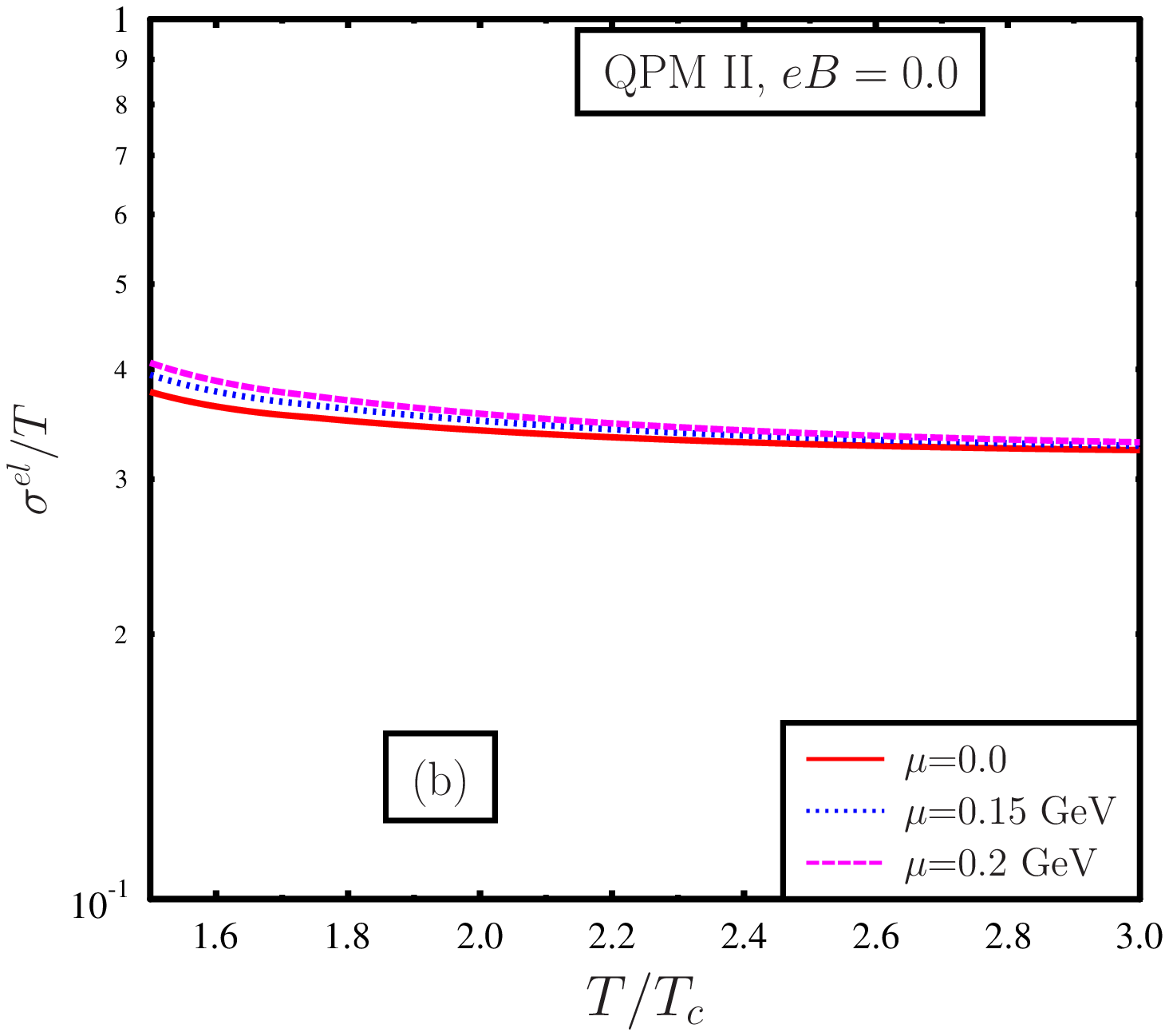}
\end{minipage}
\caption{Plot(a): Variation of normalized electrical conductivity ($\sigma^{el}/T$) with temperature for vanishing magnetic field
and different values of quark chemical potential in QPM I. Plot(b): Variation of normalized electrical conductivity ($\sigma^{el}/T$) with temperature for vanishing magnetic field
and different values of quark chemical potential in QPM II. For QPM I(plot(a)) and QPM II(plot(b)),
with increasing quark chemical normalized electrical conductivity increases.}
\label{fig4}
\end{figure}

Next we show the variation of normalized electrical conductivity with temperature for vanishing magnetic field but with finite quark 
chemical potential in Fig.\eqref{fig4}. From Fig.\eqref{fig4}(a) and Fig.\eqref{fig4}(b) we can see that for both the 
quasi particle models $\sigma^{el}/T$ increases with quark chemical potential. Variation of normalized electrical conductivity 
with quark chemical potential intimately connected with the variation of relaxation time with quark chemical potential and the 
Boltzmann factor $\exp(\pm\mu/T)$ in the equilibrium distribution function. For QPM I thermal average relaxation time has been calculated
for vanishing quark chemical potential. Hence relaxation time does not change with quark chemical potential.
At finite chemical potential number density of quarks are larger 
than antiquarks, hence in the total electrical conductivity contribution from the quarks are larger with respect to the 
antiquarks. With increasing quark chemical potential Boltzmann factor in the distribution function increases. This increasing
behaviour of the Boltzmann factor in the distribution function results in the increasing behaviour of 
normalized electrical conductivity with quark chemical potential in QPM I. On the other hand for QPM II from Fig.\eqref{fig2}(b)
we can see that with increasing quark chemical potential relaxation time decreases. But this decrease in the relaxation time 
with increasing quark chemical potential is compensated by the Boltzmann factor ($\exp(\mu/T)$) in the distribution function. Hence in QPM II
normalized electrical conductivity increases with quark chemical potential. It is also important to note that at relatively high
temperature normalized electrical conductivity does not change by a large amount with chemical potential. This is because
at high temperature Boltzmann factor ($\exp(\mu/T)$) factor is not large within the temperature and chemical potential range considered here.
\begin{figure}[h]
\centering
\begin{minipage}{0.45\textwidth}
\centering
\includegraphics[width=1.1\linewidth]{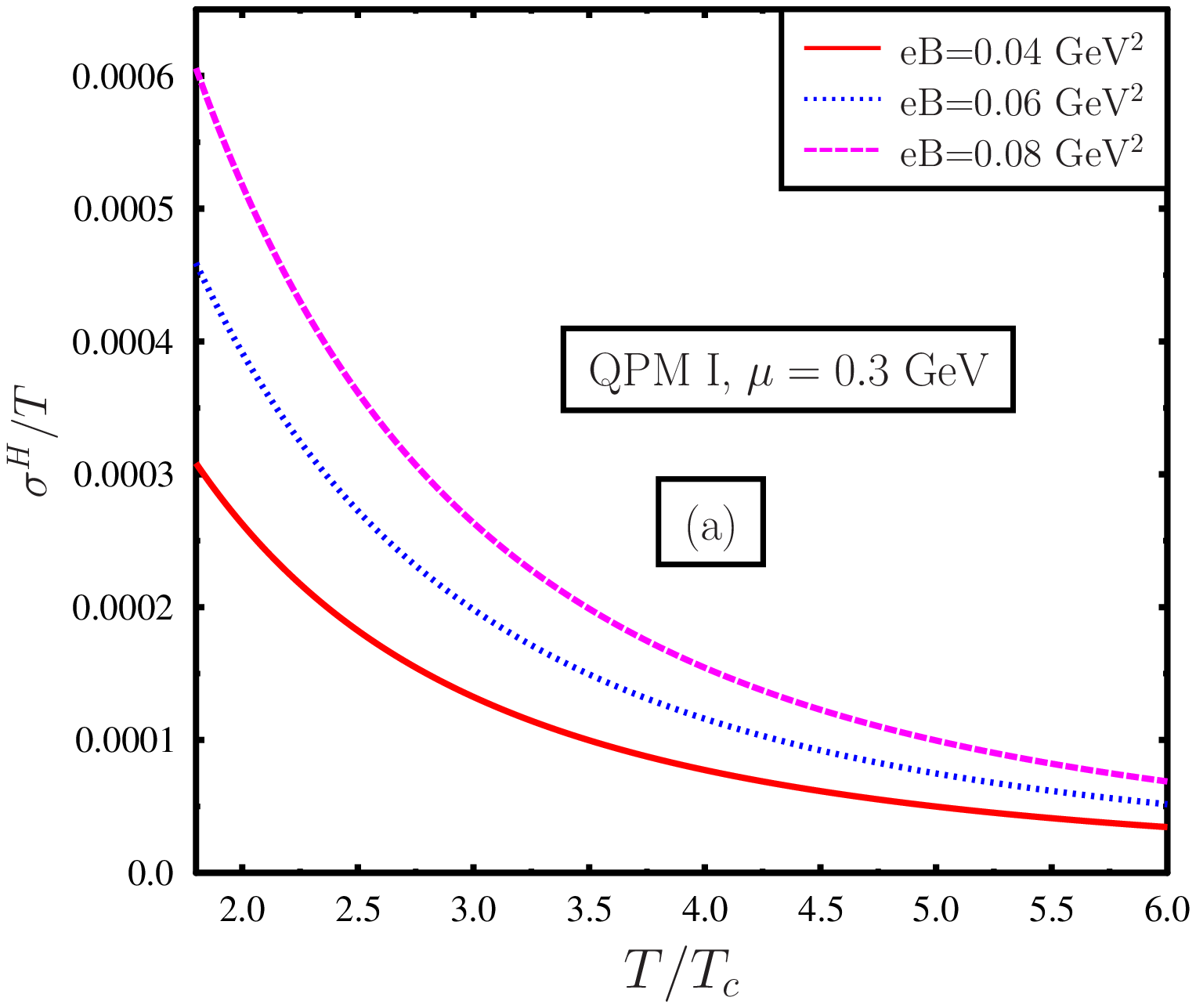}
\end{minipage}
\begin{minipage}{0.45\textwidth}
\centering
\includegraphics[width=1.1\linewidth]{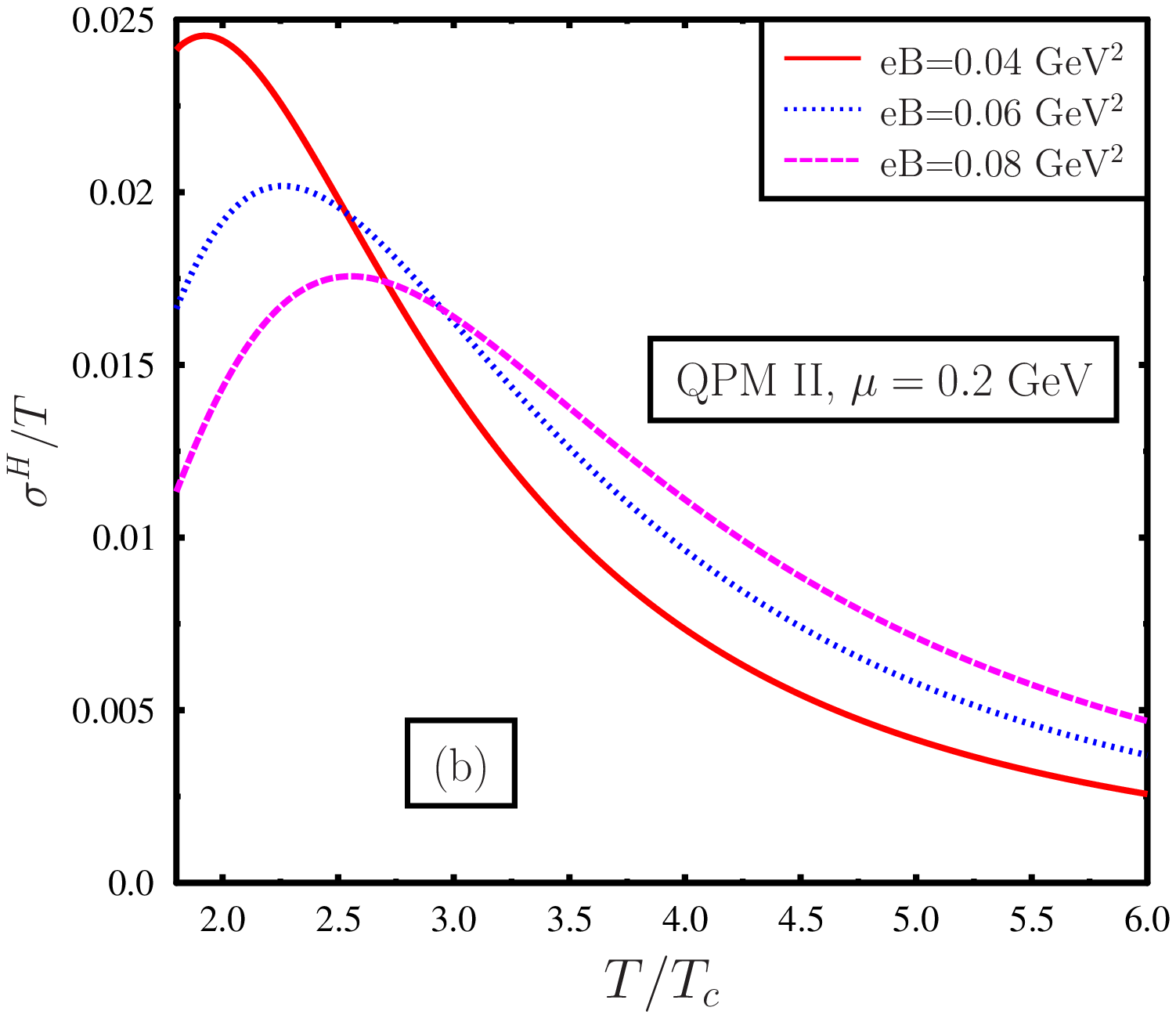}
\end{minipage}
\caption{Plot(a): Variation of normalized Hall conductivity ($\sigma^{H}/T$) with temperature for a non vanishing quark chemical
potential and various values of magnetic field in QPM I. Plot(b): Variation of normalized Hall conductivity ($\sigma^{H}/T$) with temperature for a non vanishing quark chemical
potential and various values of magnetic field in QPM II. For QPM I, as shown in plot(a),$\sigma^{H}/T$ increases with magnetic
field. Contrary to QPM I, in QPM II as shown in plot(b) for relatively low temperature normalized Hall conductivity decreases with 
magnetic field and for higher temperature it increases with magnetic field. For a fixed value of magnetic field and quark chemical
potential in QPM I (plot(a))$\sigma^{H}/T$ decreases with temperature. On the other hand 
for QPM II (plot(b)) $\sigma^{H}/T$
first increases with temperature for relatively small temperature and eventually it decreases with temperature at large temperature.}
\label{fig5}
\end{figure}

Now we turn our focus to the variation of normalized Hall conductivity $\sigma^H/T$ with temperature. For vanishing quark 
chemical potential the relaxation time of quarks and antiquarks are same. Hence in the Hall conductivity particles and antiparticle
contributions are exact but opposite. So the net Hall current at vanishing quark chemical potential is zero as can be 
explicitly seen in Eq.\eqref{equ16}. Only for non vanishing quark chemical potential Hall conductivity has non zero value. In
Fig.\eqref{fig5} we show the variation of normalized Hall conductivity ($\sigma^{H}/T$) with temperature for non vanishing
quark chemical potential and magnetic field. From Fig.\eqref{fig5}(a) for QPM I, we see that for finite quark chemical potential
Hall conductivity increases with magnetic field. However for QPM II, Hall conductivity has a non monotonic behavior with temperature,
as can be seen in Fig.\eqref{fig5}(b), where at small temperature Hall conductivity decreases with increase in magnetic field and 
at relatively high temperature Hall conductivity increases with increase in magnetic field. This different behavior of Hall 
conductivity in QPM I and QPM II is mainly due to different values of relaxation time in these quasi particle models. 
This behaviour of $\sigma^{H}/T$ as shown in  Fig.\eqref{fig5}(b) with magnetic field is plausibility due to the factor
$\frac{\omega_c\tau}{1+(\omega_c\tau)^2}$ in the Hall conductivity as can be seen from Eq.\eqref{equ16}. At relatively small 
temperature relaxation time is large and $\sigma^{H}/T\sim \frac{1}{\omega_c}$. On the other hand at high temperature relaxation
time is smaller and  $\sigma^{H}/T\sim \omega_c$. Thus at smaller temperature with increasing magnetic field normalized 
Hall conductivity decreases and at high temperature it increases with magnetic field in QPM II. 
As we have already mentioned in QPM I relaxation time is order of magnitude smaller than that of in QPM II, hence in this 
case $\sigma^{H}/T\sim \omega_c$ for the range of temperature, chemical potential and magnetic field considered here.
Hence in QPM I Hall conductivity increases with magnetic field. For a fixed magnetic field variation of normalized Hall conductivity with
temperature is rather convoluted. In QPM II, for a fixed magnetic field, at relatively low temperature due to large relaxation time,
$\sigma^{H}/T\sim f_0(1-f_0)/T^2$, which increases with increasing temperature for the parameter range considered in this work.
On the other hand at relatively high temperature, due to small relaxation time, $\sigma^{H}/T\sim \tau^2 f_0(1-f_0)/T^2$,
which decreases with increasing temperature.  
\begin{figure}[h]
\centering
\begin{minipage}{0.45\textwidth}
\centering
\includegraphics[width=1.1\linewidth]{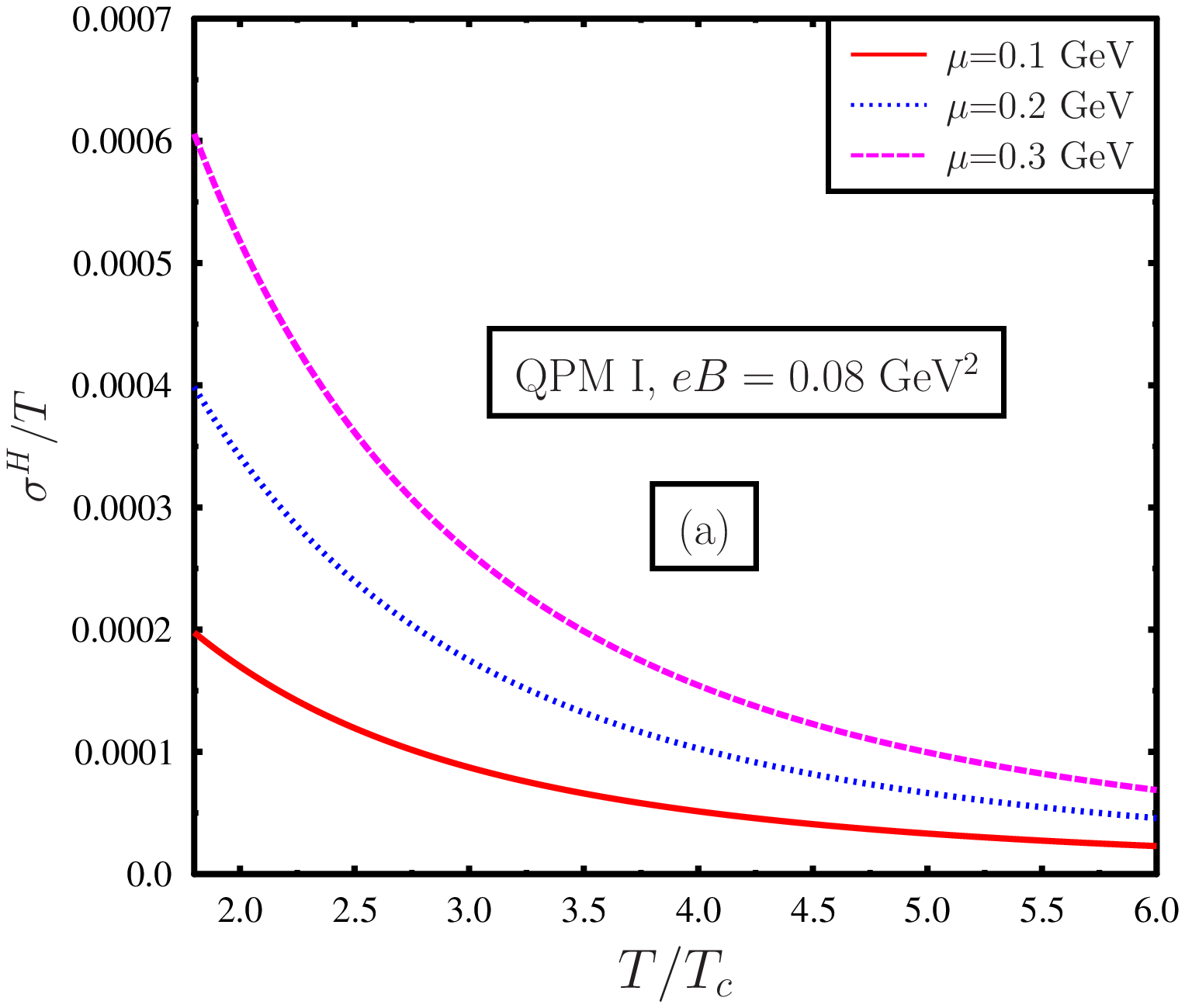}
\end{minipage}
\begin{minipage}{0.45\textwidth}
\centering
\includegraphics[width=1.1\linewidth]{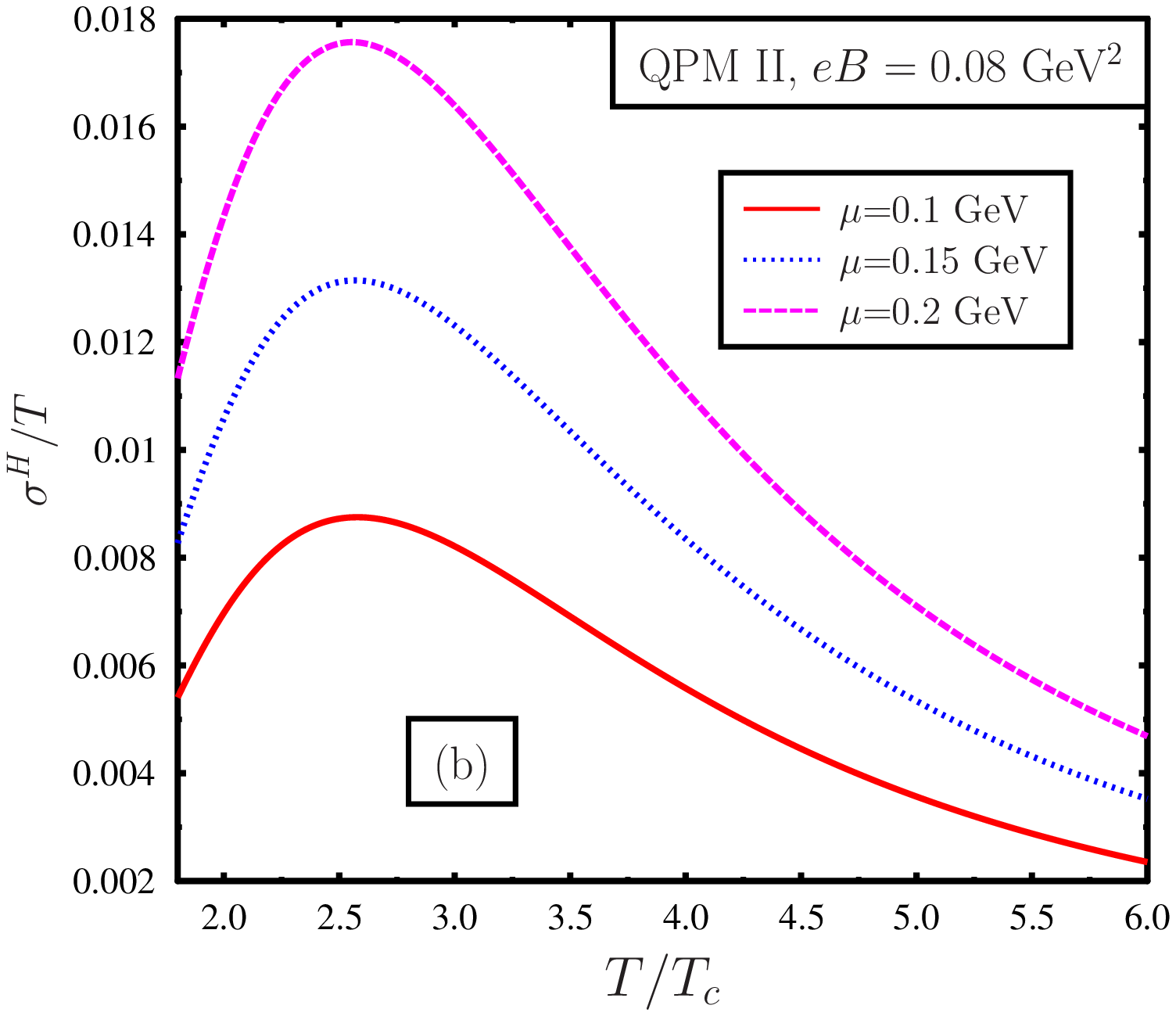}
\end{minipage}
\caption{Plot (a): Variation of normalized Hall conductivity ($\sigma^{H}/T$) with temperature ($T$) for various values of quark 
chemical potential ($\mu$) and fixed non vanishing magnetic field in QPM I. Plot (b): Variation of normalized Hall conductivity ($\sigma^{H}/T$) with temperature ($T$) for various values of quark 
chemical potential ($\mu$) and fixed non vanishing magnetic field in QPM II. From these plot we can see that with increasing
quark chemical potential $\sigma^{H}/T$ increases. For QPM I with increasing temperature normalized Hall conductivity decreases.
On the other hand for QPM II normalized Hall conductivity varies non monotonically with temperature.}
\label{fig6}
\end{figure}
 
 In Fig.\eqref{fig6} we show the variation of normalized hall conductivity ($\sigma^{H}/T$) with temperature for different
 values of quark chemical potential and finite magnetic field. For Fig.\eqref{fig6}(a) and Fig.\eqref{fig6}(b) it is clear
 that with increasing quark chemical potential Hall conductivity increasing. At vanishing quark chemical potential 
 Hall conductivity is zero because of the exact cancellation of Hall current of particles and its antiparticles. At finite 
 quark chemical potential number density of particles are larger than that of antiparticles. Hence at finite quark chemical
 potential there is a net Hall current and non vanishing Hall conductivity. With increasing quark chemical potential contribution
 of the particles increase, which result in the increasing behaviour of $\sigma^{H}/T$ with quark chemical potential as can be 
 seen in Fig.\eqref{fig6}. Similar to the Fig.\eqref{fig5} for a fixed chemical potential and magnetic field variation of 
 normalized Hall conductivity with temperature is convoluted and it depends upon various factors e.g. relaxation time, distribution function etc.
 
\section{conclusions}
In this investigation we study the electrical conductivity and Hall conductivity in the presence of magnetic field for 
quark gluon plasma within the framework of quasi particle models. Here we have considered two specific quasi particle models of 
QGP, QPM I where the quasi particle nature of the quarks and gluon are encoded in thermal mass of the quasi partons and 
QPM II, where the quasi particle nature is encoded in the effective fugacity parameter in the distribution function. Both of these
models have been explored earlier in literature. We found that with increasing magnetic field normalized electrical conductivity
($\sigma^{el}/T$) decreases in both quasi particle models. However due to widely different values of relaxation time in these 
models, decrease in normalized electrical conductivity with magnetic field is larger in QPM II than its counterpart in QPM I. In 
 both these models thermal averaged relaxation time decreases with temperature. In QPM II relaxation time 
decreases with increasing quark chemical potential, plausibility due to increasing number density of quasi partons and increasing
interaction rates. In both models normalized electrical conductivity increases with increasing quark chemical potential.
For a pair plasma at vanishing quark chemical potential Hall current due to particles and antiparticles cancels each other.
Hence Hall conductivity can have a non zero value only for finite quark chemical potential. In QPM I with increasing magnetic field normalized Hall conductivity ($\sigma^{H}/T$) increases. On the other hand for QPM II, at relatively small
temperature, $\sigma^{H}/T$ decreases with increasing magnetic field and at high temperature $\sigma^{H}/T$
increases with increase in magnetic field. Finally with finite magnetic field $\sigma^{H}/T$ increases with increase in quark chemical potential. This is due to the fact that at finite quark chemical
potential contributions of the particles are larger than their antiparticles in the net Hall conductivity and with increasing
chemical potential contribution of the quarks increases, resulting in an increasing behaviour of $\sigma^{H}/T$ with quark chemical potential.

\section*{Acknowledgement}
RKM would like to thank Theoretical Physics Division of Physical Research Laboratory, Ahmedabad
for support and local hospitality for her visit,
during which this problem was initiated.
Also, RKM would like to thank Basanta K. Nandi and Sadhana Dash for constant support and encouragement.

\end{document}